\pdfoutput=1
\documentclass[reprint,amsmath,aps,prb,footinbib]{revtex4-2}
\usepackage{graphicx,epstopdf,dcolumn,bm,mathtools,siunitx,multirow,hyperref,xcolor,booktabs,physics}

\begin{document}
\title{Polaron catastrophe within quantum acoustics}

\author{Alhun~Aydin$^{1,2}$, Joonas~Keski-Rahkonen$^{2,3}$, Anton M.~Graf$^{2,3,4}$, Shaobing~Yuan$^{5}$, Xiao-Yu~Ouyang$^{6}$, \"{O}zg\"{u}r E. M\"{u}stecapl{\i}o\u{g}lu$^{7,8}$, Eric J.~Heller$^{2,3}$}
\affiliation{$^1$Faculty of Engineering and Natural Sciences, Sabanci University, 34956 Tuzla, Istanbul, T\"urk{\.i}ye}
\affiliation{$^2$Department of Physics, Harvard University, Cambridge, Massachusetts 02138, USA}
\affiliation{$^3$Department of Chemistry and Chemical Biology, Harvard University, Cambridge,
Massachusetts 02138, USA}
\affiliation{$^4$Harvard John A. Paulson School of Engineering and Applied Sciences,
Harvard, Cambridge, Massachusetts 02138, USA}
\affiliation{$^5$Department of Chemistry, Princeton University, Princeton, NJ 08544, USA}
\affiliation{$^6$Division of Chemistry and Chemical Engineering, California Institute of Technology, Pasadena, CA 91125, USA}
\affiliation{$^7$Department of Physics, Ko{\c{c}} University, 34450 Sar{\i}yer, Istanbul, T\"{u}rkiye}
\affiliation{$^8$T\"{U}B\.{I}TAK Research Institute for Fundamental Sciences, 41470 Gebze, T\"{u}rkiye}

\date{\today}
\begin{abstract}
The quantum acoustic framework has recently emerged as a non-perturbative, coherent approach to electron-lattice interactions, uncovering rich physics often obscured by perturbative methods with incoherent scattering events. Here, we model the strongly coupled dynamics of electrons and acoustic lattice vibrations within this framework, representing lattice vibrations as coherent states and electrons as quantum wavepackets, in a manner distinctively different from tight-binding or discrete hopping-based approaches. We derive and numerically implement electron backaction on the lattice, providing both visual and quantitative insights into electron wavepacket evolution and the formation of acoustic polarons. We investigate polaron binding energies across varying material parameters and compute key observables—including mean square displacement, kinetic energy, potential energy, and vibrational energy—over time. Our findings reveal the conditions that favor polaron formation, which is enhanced by low temperatures, high deformation potential constants, slow sound velocities, and high effective masses. Additionally, we explore the impact of external electric and magnetic fields, showing that while polaron formation remains robust under moderate fields, it is weakly suppressed at higher field strengths. These results deepen our understanding of polaron dynamics and pave the way for future studies into non-trivial transport behavior in quantum materials.
\end{abstract}
\maketitle
\section{Introduction}
Understanding the interplay between charge carriers and lattice vibrations is a cornerstone mission in condensed matter physics, holding implications for a wide range of phenomena from electrical and thermal transport to high-temperature superconductivity~\cite{Many-Particle,Keimer2015}. Of late, by treating lattice vibrations as wave-like entities with respect to coherent states (rather than in terms of Fock states emphasizing the particle nature), and charge carriers as quantum wavepackets, we have cast the problem as time-dependent in real space and subsequently established a non-perturbative and coherent treatment of electron-phonon interaction~\cite{Heller22}. 

While coherent states have been employed to describe ensembles of phonons in earlier studies~\cite{noolandi_use_1972,van_kranendonk,ichikawa1976coherent,vzmuidzinas1978electron,PhysRevB.65.174303}, and in some more recent ones~\cite{PhysRevB.72.014307,grusdt2024impurities,PhysRevLett.124.223401,christianen2024phase}, including the extensions of the Gaussian ansatz for phonons with multi-mode squeezing terms~\cite{shi2018variational}, and the wavepacket propagation approaches have been implemented~\cite{vzmuidzinas1978electron,davydov1986quantum,larson2007dynamics}, these efforts were limited to specific systems and lacked the computational advancements needed to harness the full efficiency of such approaches. By contrast, we have put forward a fresh framework for the electron-lattice interaction~\cite{Heller22}, making clear that the conventional perturbative treatment of electrons and lattice vibrations leads to approximations obscuring the hidden physics within the standard Fröhlich and Holstein models~\cite{Heller22,Aydin24,HellerDDP24,e26070552}. This program, in parallel to quantum optics~\cite{PhysRevLett.10.277,glauber1963} -- called \emph{quantum acoustics} -- has led to rapid advances. It not only delivered the missing dual to the number state representation of lattice vibrations, recovering the conventional results in the perturbative limit~\cite{Heller22}, but also provided deep insight and answers into several longstanding enigmas of strange metals.  This includes the origins of the paradigmatic T-linear resistivity, Mott-Ioffe-Regel violation, and displaced Drude peak phenomena~\cite{Aydin24,HellerDDP24}.

Building on this progress, we demonstrate here that the quantum-acoustic picture with the deformation potential approach provides a versatile tool to investigate the intricate dialog between the lattice and electrons that give rise to polarons. Another form of electron-phonon coupling, the Peierls (Su-Schrieffer-Heeger) interaction, modulates hopping amplitudes rather than on-site potentials, leading to distinct polaron properties such as self-trapping transitions, light bipolarons, and possible links to high-Tc superconductivity~\cite{PhysRevB.35.4291,PhysRevB.35.4297,sous2017phonon,PhysRevLett.121.247001,PhysRevB.102.235145,PhysRevX.13.011010}. While considerable attention has been paid to momentum-space dynamics, the real-space and real-time behavior of polarons have received less attention (for a rare exception, see Ref.~\cite{PhysRevB.83.075104}). Real-time studies using variational or Green’s function cluster expansion techniques exist~\cite{shi2018variational,li2019variational,PhysRevB.104.035106}, but they are not performed in position space. Furthermore, most research to date has focused on the optical branch of lattice vibrations~\cite{polaronreview,Devreese_2009}, but another kind of a polaron can form when an electron couples to longitudinal acoustic lattice vibrations. This aspect of acoustic polarons was first explored by Schüttler and Holstein in 1983~\cite{PhysRevLett.51.2337,SCHUTTLER198693}, and later studies have examined the stability~\cite{doi:10.1143/JPSJ.70.2968,PhysRevB.15.4822,SUN2022414360}, mobility~\cite{KORNJACA2018183}, lifetime~\cite{IVIC2002144}, and self-trapping transitions~\cite{PhysRevB.32.3515, PhysRevB.62.3241, PhysRevB.86.144304} of acoustic polarons for varying electron-phonon coupling strengths, as well as the exciton-phonon coupling~\cite{doi:10.1126/science.adf2698,shih24,PhysRevB.109.045202} and ground-state properties~\cite{PhysRevB.34.5912,MANFOUO2022414172,PhysRevB.54.12835,PhysRevB.104.L161111}. In the context of ultracold atoms, Bose polarons have been extensively studied using the Fröhlich model, where the role of lattice vibrations is played by Bogoliubov phonons of a Bose-Einstein condensate, and the coherent-state representation has been employed to investigate their equilibrium and dynamical properties, particularly at $T=0$K, with extensions to finite temperature also considered~\cite{skou2021non,PhysRevA.105.053302,grusdt2024impurities,PhysRevLett.124.223401}. More recently, acoustic polarons have been investigated utilizing diagrammatic quantum Monte Carlo techniques~\cite{PhysRevLett.102.186801,Granger2012,PhysRevB.104.L161111}, as well as density functional theory with variational wavefunctions~\cite{shih2024theory,britt2024momentum}, and the dynamics in the strong-coupling limit~\cite{griesemer2017dynamics}. In contrast to these studies, the quantum-acoustic perspective embraced here opens up an unexplored pathway to directly observe the elusive birth and life of acoustic polarons in space over time.

In our previous studies taking advantage of the quantum-acoustic approach~\cite{Aydin24,HellerDDP24, zimmermann_entropy_2024}, we have been restricted to temperatures above $50\, \textrm{K}$, where the electron feedback onto the lattice motion can be justifiably omitted. However, as the temperature decreases, it becomes increasingly important to consider the electron feedback upon lattice dynamics. In fact, the strong back-action of electrons to the underlying lattice can sire a polaron~\cite{Emin_2012}. 

Here, we present the full quantum-acoustical treatment for the venerable Fröhlich model, and subsequently demonstrate that electrons can form acoustic polarons by interacting non-perturbatively and coherently with lattice vibrations, represented by coherent states. In an approach inaccessible to Fock state strategies, we find the emergence of a polaron in the midst of itinerant potential energy hills and valleys stemming from thermal lattice vibrations. The polarons stabilize themselves by shedding their energy of formation as lattice waves, seen below as if a rock had been thrown into a pond. We show that these acoustic polarons remain robust across variations in system parameters, making them a reliable and stable phenomenon at low temperatures. Finally, we discuss the implications of our results and the coherent state formalism of lattice vibrations in general and conclude with a brief summary.

\section{Quantum-acoustical model}\label{s:sec2}

\begin{figure*}
\centering
  \includegraphics[width=0.99\textwidth]{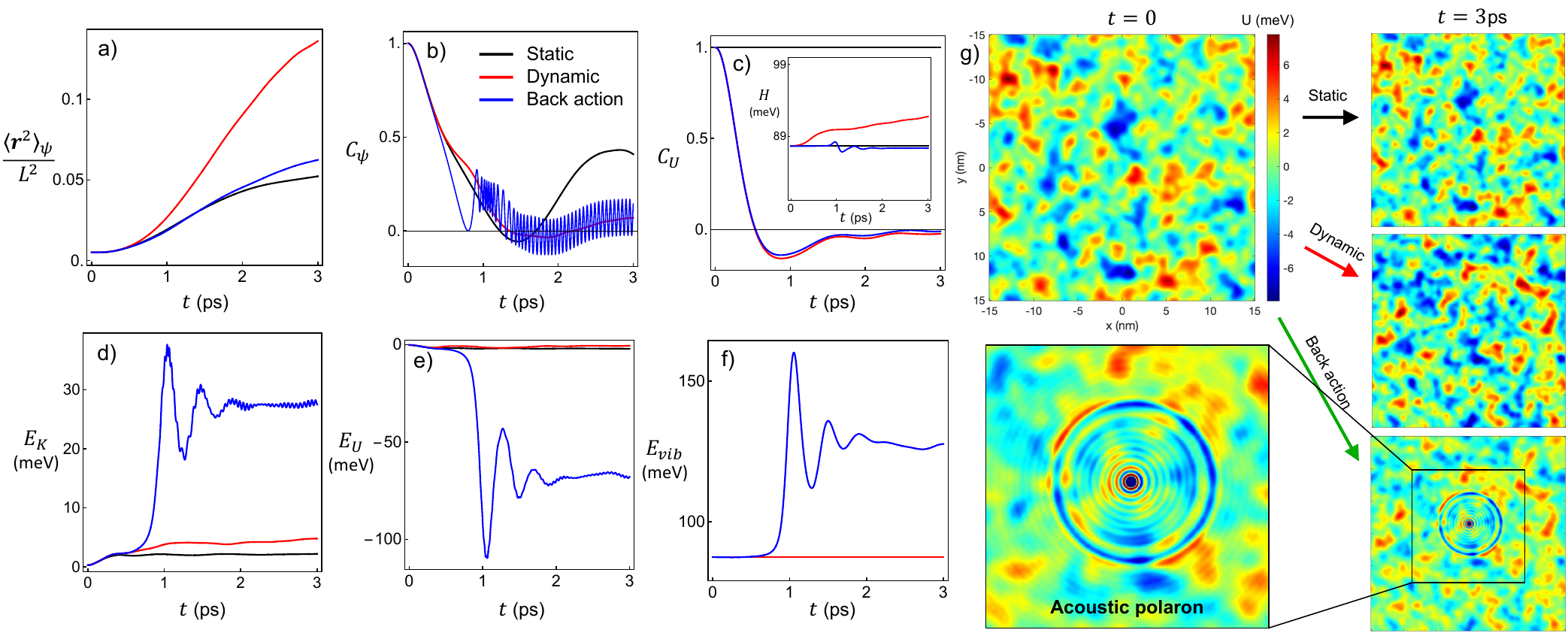}
  \caption{Comparison of observables under static (black) and dynamic (red) deformation potential, and deformation potential with back action (blue), with no external fields applied. Time variations of a) mean square distance, b) time-autocorrelation of the electron wavefunction, c) time-autocorrelation of the potential, where the time variation of the total energy is given as an inset. The expectation values of d) electron kinetic energy, e) potential energy, f) vibrational energy, are shown. g) The deformation potential at $t=0$ is shown alongside snapshots at $t=3$ps for static, dynamic, and back action cases in the right column. The self-trapping effect (a strongly bound electron-lattice state) becomes evident when the electron’s back action on the lattice is included. The zoomed-in view demonstrates the formation of an acoustic polaron in real space, with a ripple effect in the lattice coming from the energy released from the polaron. 
  }\label{fig1}
\end{figure*}

Within the coherent state paradigm embracing the wave nature, we describe an electron by a quantum wavepacket and the lattice vibrations as a quantum field, in lieu of phonons. In fact, we eschew the word ``phonons'' in our context because it refers to a quantized particle; Instead, we treat the lattice in a similar manner as the classical electromagnetic field limit of quantum optics. Employing the deformation potential approach and transitioning into real space confronts us with a disordered and dynamic energy landscape attributable to thermal lattice vibrations. The electrons quasi-elastically deflect from the bumps and dips of this ever-morphing potential, just as Bardeen and Shockley stated in 1950~\cite{BS2}: 
\begin{quote}
     This implies that the pertinent acoustical waves can be treated by classical methods, even at fairly low temperatures.
\end{quote}
As characterized in Ref.~\cite{BS2}. most electron deflections are then elastic, although today we might rather define them as quasielastic. But certainly not always inelastic, as the Bloch-Gr\"uneisen theory demands. The perturbative approach was subsequently born of necessity at the time. Unfortunately, this interaction has ever since been unilaterally treated by quantized perturbation methods and Boltzmann scattering, requiring an inelastic event for every deflection.

More than 70 years later, we have restored the original program~\cite{Heller22}. In particular, we have returned back to the standard electron-phonon Hamiltonian in second-quantized form~\cite{Many-Particle};
\begin{eqnarray}
\begin{aligned}
\hat{H} &=\hat{H}_{\text{e}}+\hat{H}_{\text{ph}}+\hat{H}_{\text{e-ph}} \\
&=\sum_\mathbf{k}\epsilon_\mathbf{k} c_\mathbf{k}^{\dagger} c_\mathbf{k} + \sum_\mathbf{q} \hbar\omega_\mathbf{q}(a_\mathbf{q}^{\dagger}a_\mathbf{q}) \\
&+ \sum_{\mathbf{k},\mathbf{q}}g_{\mathbf{k},\mathbf{q}}c_{\mathbf{k}+\mathbf{q}}^{\dagger} c_\mathbf{k}(a_\mathbf{q}+a^{\dagger}_{-\mathbf{q}}),
\end{aligned}
\end{eqnarray}
where $\epsilon_\mathbf{k}$ is electron band energy with $\mathbf{k}$ wavenumber, $c_\mathbf{k}^{\dagger}$ ($c_\mathbf{k}$) is creation (annihilation) operators for electrons, $\omega_\mathbf{q}$ is the frequency of the phonon normal mode $\mathbf{q}$, $a_\mathbf{q}^{\dagger}$ ($a_\mathbf{q}$) is creation (annihilation) operators for phonons, and $g_{\mathbf{k},\mathbf{q}}$ is electron-phonon coupling strength. In the Hamiltonian, we have omitted the zero-point energy of the lattice, as it constitutes a constant offset that does not influence the dynamics or interaction terms relevant to the analysis of polaron formation.

Following the path forged by quantum optics~\cite{scully1997quantum, walls2007quantum}, we figure the evolution of the deformation potential composed of coherent states $\vert \alpha_{\mathbf{q}} \rangle$ in terms of the expectation values $\alpha_{\mathbf{q}} = \langle \alpha_{\mathbf{q}} \vert a_{\mathbf{q}} \vert \alpha_{\mathbf{q}} \rangle$, instead of considering the time-evolution of the field operators. We initialize each coherent state as
\begin{align}
\alpha_\mathbf{q}(t_0)=\sqrt{\langle n_{\mathbf{q}}\rangle_{\text{th}}}e^{i\phi_\mathbf{q}},
\end{align}
with the random phase $\phi_{\mathbf{q}}$ and thermal amplitude $\langle n_{\mathbf{q}}\rangle_{\text{th}}=[\exp\left(\hbar\omega_{\mathbf{q}}/k_{\textrm{B}} T\right) - 1]^{-1}$. 

The coherent state approach presented here serves as the dual counterpart to the traditional number state description of electron–lattice dynamics in principle, but it leads to a vastly different set of possible approximations. As blueprinted in Ref.~\cite{Heller22}, each normal mode of lattice vibration with wavevector $\mathbf{q}$ is fundamentally associated with a corresponding coherent state $\vert \alpha_{\mathbf{q}} \rangle$. Furthermore, at thermal equilibrium, each mode is treated as being in contact with a heat bath at temperature $T$, resulting in thermal ensembles of coherent states where the average occupation of each mode follows the Bose–Einstein distribution. By taking into account the independence of normal modes, the collective lattice vibration $\vert \chi \rangle$ can be described as the product of coherent states of the individual normal modes -- essentially a multimode coherent state $\vert \chi \rangle = \otimes_{\mathbf{k}} \vert \alpha_{\mathbf{k}} \rangle$, as minutely discussed in Ref.~\cite{hellerkim}.

We initialized the charge carrier by the following Gaussian wave packet,
\begin{equation}\label{Initial_Gaussian}
    \psi(\mathbf{r}, 0) = \mathcal{N}\exp\left(-\frac{1}{4} \vert \mathbf{r} \cdot \boldsymbol{\sigma}^{-1} \vert^2  - i \mathbf{k} \cdot \mathbf{r} \right),
\end{equation}
where $\mathcal{N}$ is the normalization factor, $\mathbf{r}$ is the position vector, $\boldsymbol{\sigma}^{-1} = (\sigma_x^{-1}, \sigma_y^{-1})$ describing the initial width of the wavepacket. Without loss of generality, we can choose to launch the test wavepacket into the $x$ direction with the initial momentum $\hbar k$, thus $\mathbf{k} = (k, 0)$, and assume equal widths in both spatial directions.
 
Motivated by Refs.~\cite{leopold2021derivation, leopold2021landau, leopold2022landau}, we consider the evolution of an initial state $\vert \Psi \rangle$ of product form
\begin{equation}\label{Eq:LP_ansatz}
    \vert \Psi \rangle  = \vert \psi \rangle \otimes \vert \chi \rangle
\end{equation}
combining the electronic state $\vert \psi \rangle$ and the lattice state $\vert \chi \rangle$. This starting point is equivalent to employing the time-dependent Hartree approximation. However, due to the mutual interaction, the system will build-up correlations between the electron $\vert \psi \rangle$ and the coherent state field $\vert \chi \rangle$ and the solution of the Schr{\"o}dinger equation $i\hbar \partial_t \vert \Psi \rangle = \hat{H} \vert \Psi \rangle$ will no longer be of product form, but it can be well approximated by a product state as in Eq.~\ref{Eq:LP_ansatz} obeying the Landau–Pekar-like equations (for historical reference, see, e.g. Refs.~\cite{pekar1946local, landau1948effective}). While Lee-Low-Pines (LLP)~\cite{PhysRev.90.297} could have been useful at $T=0$K, here we consider finite temperatures, where lattice vibrations are thermal and in the semiclassical regime. Since we assume an initially uncorrelated electron-phonon system, applying the LLP transformation first would not provide a significant advantage in this setting.

In this scheme, the electron and coherent states defining the lattice field evolve dynamically through the following coupled, nonlinear equations of motion,
\begin{subequations}
\begin{align}\label{coupled}
    i\hbar \frac{\partial \psi}{\partial t} & = \left[\frac{1}{2m^{*}} (i\hbar\nabla + e\mathbf{A})^2  + e\varphi +  V_D(\mathbf{r},t) \right]\psi, \\
    i\hbar\frac{\partial\alpha_\mathbf{q}}{\partial t} & = \hbar\omega_\mathbf{q}\alpha_\mathbf{q} + g_\mathbf{q} \int e^{-i\mathbf{q}\cdot\mathbf{r}}|\psi(\mathbf{r}, t)|^2\, \textrm{d}\mathbf{r}.
\end{align}
\end{subequations}
along with the vector $\mathbf{A}$ and scalar potential $\varphi$ stemming from an static, external magnetic $\mathbf{B} = \nabla \times \mathbf{A}$ and electric field $\mathbf{E} = -\nabla\varphi$. Here, we use the effective mass approximation for the electron, where $m^{*}$ denotes its effective mass. The quasi-classical lattice field in real space is given by,
\begin{eqnarray}\label{quasiclassical_DP}
\begin{aligned}
    V_D(\mathbf{r},t) & = \langle \chi \vert \sum_{\mathbf{q}} g_\mathbf{q} (a_\mathbf{q}+a^{\dagger}_{-\mathbf{q}}) \vert \chi \rangle \\
    & = 2\mathrm{Re}\left[\sum_{\mathbf{q}}g_\mathbf{q}\alpha_\mathbf{q}e^{i\mathbf{q}\cdot\mathbf{r}}\right],
\end{aligned}
\end{eqnarray}
where the Fröhlich-type~\cite{frohlich1954electrons} coupling constant is expressed as $E_d\sqrt{\hbar/(2\rho\mathcal{V}\omega_\mathbf{q})}|\mathbf{q}|$, consisting the wave vector $\mathbf{q}$ of a normal mode, the deformation potential coupling constant $E_d$, the mode frequency $\omega_{\mathbf{q}}$, the mass density $\rho$, and the volume (area in two dimensions) $\mathcal{V}$ of the lattice. The deformation potential constant, $E_d$, characterizes the strength of the electron-phonon interaction arising from the modulation of the electronic energy due to lattice vibrations. It is a material-dependent parameter typically extracted from experiments or ab initio calculations. We neglect the explicit $\mathbf{k}$-dependence of the electron-phonon coupling, focusing on the dominant $\mathbf{q}$-dependent deformation potential, which suffices to capture the essential polaron dynamics. Moreover, we employ the Debye model, introducing the linear dispersion $\omega_{\mathbf{q}}=v_s|\mathbf{q}|$ where $v_s$ is sound speed, and restricting to the wavevector summation to Debye wavevector due to the minimal lattice spacing.

The deformation potential so defined aligns with the earlier works of Refs.~\cite{Heller22, Aydin24}, where the same deformation potential arises from the displacement field argumentation of the longitudinal acoustic modes of the lattice vibrations. In direct analogy to quantum optics~\cite{scully1997quantum, walls2007quantum}, the coherent state representation leads from the quantized lattice vibrations naturally to a quasi-classical field as “real” as any external field, acting on the electrons. The quasi-classical coherent states $\alpha_{\mathbf{q}}$ behave like driven harmonic oscillators, where an explicit force is associated with the electron density acting upon the lattice.

As studied in Ref.~\cite{Aydin24}, the electron response to the lattice field is almost of no consequence if all the electron-lattice couplings related to the allowed lattice modes are small compared to the corresponding mode energies. The time-dependence of the deformation potential is then governed by the wave equation. Therefore, the acoustic lattice disorder field appears as a chaotic sea of roaming sound waves, which has a close resemblance to the vector potential of the blackbody field by Hanbury Brown and Twiss~\cite{HBT1, HBT2}, except for the existence of the ultraviolet cutoff determined by the Debye wavevector. Notice that our system exhibits the characteristics of Fröhlich polarons due to its long-range, momentum-dependent electron-phonon interaction (deformation potential coupling), unlike Holstein polarons, which feature purely local on-site coupling and are typically studied on discrete lattices.

The deformation potential is itself a peculiar entity: it is homogeneously random~\cite{ziman1979models}, translating that the potential  $V_D(\mathbf{r}, t)$ averages to zero over space, or over time at a given spot (with the assumption of random phases). Moreover, the potential is statistically the same if inverted. This special type of spatial-temporal correlation sets the deformation potential apart from other types of lattice distortions, for instance, ones commonly investigated in the context of Anderson localization~\cite{Anderson_localization}. Furthermore, this feature implies that the electrons are itinerant because any valley they settle into is about to become a hilltop. The equations we have presented tell of a dynamical relationship between electrons and this unruly but deformable underlying lattice; a perturbative approach would be entirely inadequate. 

In the study below, we focus on the case of a single electron within the effective mass description, which wanders through the dynamical but malleable landscape of hills and valleys of the lattice potential.  As a first step, we can look for a steady state solution for the electron-lattice equations. Without the external fields, the steady state solution is determined by a form of Gross-Pitaevskii equation:
\begin{align}\label{eq: nlshrodinger}
    E\psi(\mathbf{r})= \left(-\frac{\hbar^2}{2m}\nabla^2-\frac{E_d^2}{\rho v_s^2}|\psi(\mathbf{r})|^2\right)\psi(\mathbf{r}),
\end{align}
which is known to support soliton solutions (see, e.g., Ref.~\cite{dalfovo1999theory}). In other words, a stable and localized wavepacket is possible due to the nonlinear term favoring spatially confined electron density. This feature already implies the possible existence of polarons, further exemplified by a similar approach exploited for investigating polarons in two-dimensional atomic crystals~\cite{sio2023polarons}. Moreover, the existence of a polaron within the Fröhlich model is known from many works, varying from the strong~\cite{pekar1946local, Evrard1965OnTE, devreese1964excited, lieb1977existence} to the weak coupling limit~\cite{frohlich1954electrons} as well as the all-coupling theories~\cite{feynman1955slow, mishchenko2000diagrammatic}. However, all these considerations are agnostic as to the dynamical aspect of the polaron formation.

To address this shortfall, we turn to quantum wavepacket propagation techniques that are virtually demanded by the time-varying deformation potential. More specifically, our investigation of electron dynamics under the defined effective equation of motion approaches the issue from the point of view of Gaussian wavepackets that are a common tool for analyzing time-dependent features of a quantum system~\cite{heller2018semiclassical, tannor2007introduction}, for instance in the studies of quantum optics~\cite{scully1997quantum, walls2007quantum}, scarring~\cite{keski2019quantum, PhysRevB.96.094204.2017, keski2024antiscarring}, and branched flow~\cite{heller2021branched, superwire1, superwire2}.

For the propagation of the wavepacket, we utilize the (third-order) split operator method~\cite{SplitOperator, Feit1982, tannor2007introduction, heller2018semiclassical}. Interestingly, this method can be extended to include a perpendicular, homogeneous magnetic field with no additional approximations, and without compromising the original proficiency, thanks to the exact factorization of the kinetic energy presented in Refs.~\cite{aichinger_comput.phys.commun_171_197_2005, janecek_phys.rev.B_77_245115_2008}.  

In general, the memory of the initial form of the wavepacket is rapidly scrambled by the chaotic disorder landscape, and its exact form is thus unimportant. Nevertheless, the electron wave, originally a Gaussian, evolving under the influence of the dynamic lattice wave field, converting the always-accessible wave nature of lattice vibrations into something valuable, a point where quantum-acoustical perspective becomes tangible. In contrast to previous studies~\cite{Heller22, Aydin24, HellerDDP24, zimmermann_entropy_2024}, we here see the effect of the electron to the lattice taking in place at low temperatures, and cannot be justifiably ignored anymore. As demonstrated and analyzed below, this electron back-action leads to the birth of acoustic polarons, as an electron creates its own trench within the potential landscape where it then settles in.

\section{Acoustic polaron formation}\label{s:sec3}

For our quantum-acoustical simulations, we choose a reference material based on hole-doped cuprates, due to their low Fermi energy and strong deformation potential coupling, which drive electron-lattice scattering into the coherent and non-perturbative regime~\cite{Lanzara2001,Heller22}. The material parameters used are: lattice parameter $a=\SI{0.38}{nm}$~\cite{PhysRevLett.89.107001}, interplanar spacing $d=\SI{1.1}{nm}$, sound velocity $v_s=\SI{4000}{m.s^{-1}}$~\cite{PhysRevB.52.R13134,SMconstants,PhysRevB.100.241114}, deformation potential constant $E_d=\SI{10}{eV}$, mass density $\rho=\SI{7000}{kg.m^{-3}}$, and effective mass $m^{*}=10m_e$~\cite{PhysRevB.72.060511,Legros2018}, where $m_e$ is the bare electron mass. We launch a Gaussian wavepacket with some initial velocity in the deformation potential generated by the acoustic lattice vibrations. Based on our calculations, the acoustic polaron formation occurs at temperatures below 20~K for the chosen parameters. Therefore, we set $T=10$ K in our simulations. The simulation frame and the initial spread of the wavepacket are chosen to be $L=L_x=L_y=\SI{50}{nm}$ and $\sigma_x=\sigma_y=0.05L$ respectively. The electron wavepacket interacts with the underlying deformation potential for a duration on the order of picoseconds, and we calculate the time variation of observables during the simulation.

\begin{figure}
\centering
  \includegraphics[width=0.49\textwidth]{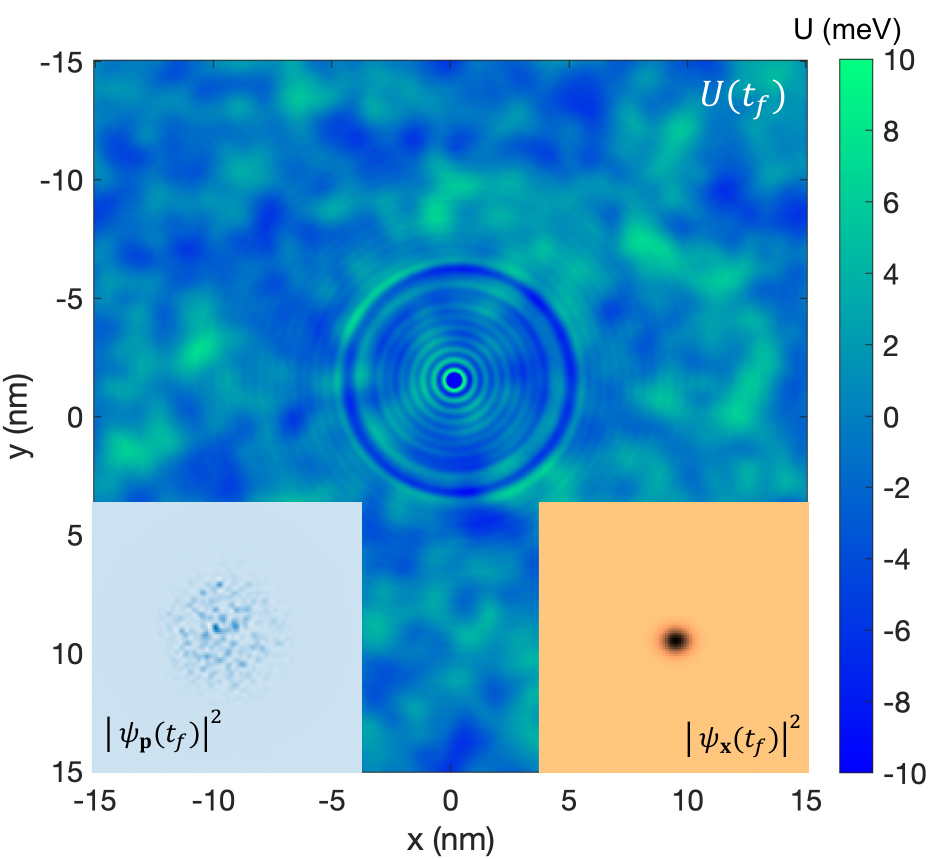}
  \caption{Snapshot of an acoustic polaron formation in real space at $T=10$ K. The main figure displays the emergent deformation potential field due to the back action of the electron onto the lattice. Left (right)-bottom corner insets show the momentum (position) space probability distributions for the electron. The fairly sudden polaron formation and local energy lowering send out circular deformation waves at the sound speed, seen at some time after formation in each of the snapshots.}\label{fig2}
\end{figure}

The average position in the x-direction is calculated as $\ev{\mathbf{x}}_\psi(t)=\bar{\mathbf{x}}=\mel{\psi(t)}{\mathbf{x}}{\psi(t)}$, and the mean square distance as $\ev{\mathbf{r}^2}_\psi(t)=\mel{\psi(t)}{[(\mathbf{x}-\bar{\mathbf{x}})^2+(\mathbf{y}-\bar{\mathbf{y}})^2]}{\psi(t)}$. The kinetic energy of the electron is given by $E_{K}(t) = \langle \psi_{\mathbf{p}}(t) | \mathbf{K} | \psi_{\mathbf{p}}(t) \rangle$, where $\psi_{\mathbf{p}}$ is the momentum space electron wavefunction. The potential energy, described by the deformation potential including the back action of the electron, is calculated as $E_{U}(t) = \langle \psi_{\mathbf{x}}(t) | U | \psi_{\mathbf{x}}(t) \rangle$, where $\psi_{\mathbf{x}}$ is the position space electron wavefunction. The energy of the lattice vibrations is given by $E_{\text{vib}}(t) = \sum_{\mathbf{q}} \hbar\omega_{\mathbf{q}} |\alpha_{\mathbf{q}}(t)|^2$. The total energy is given by $H=E_K+E_U+E_{\text{vib}}$. The autocorrelation functions of the wavepacket and the potential are calculated as $C_\psi(t)=\text{Re}[\langle \psi(0) | \psi(t) \rangle]$, and $C_U(t)=\langle U(t) | U(0) \rangle/\langle U(0) | U(0) \rangle$, respectively.

We begin by exploring polaron formation in the absence of external electric and magnetic potentials, $\varphi=0$, $\mathbf{A}=0$. In Fig. \ref{fig1}, we compare the results of the frozen deformation potential (static), time-dependent deformation potential without back action (dynamic), and the dynamic deformation potential with the electron affecting the lattice vibrations (back action). Under the frozen deformation potential, the spatial disorder stemming from thermal lattice vibrations results in the Anderson localization of the electron wave packet~\cite{PhysRevLett.47.882,Heller22,e26070552}, which is evident from the saturation of the mean square distance. When the time dependence is on (dynamic case), continuously morphing hills and valleys of the deformation potential disrupt the localizing interference patterns, leading to a transient localization, where the wave packet can extend further in space, Fig. \ref{fig1}a, before reaching the simulation boundaries after around 1 ps. \cite{Fratini, afmTL, PhysRevLett.132.266502, Aydin24}.

With the inclusion of the electron’s back action on the lattice, the electron becomes trapped by non-perturbative interactions with the lattice vibrations, resulting in the formation of an acoustic polaron. This process is visually evident in the real-space images shown in Fig. \ref{fig1}(g), where the electron digs a hole in the potential. As the hole is formed, the electron and lattice find a lower energy solution and release the energy as acoustic waves, revealed in the figure as if a pebble were thrown into a pond. This process, which we call the polaron catastrophe, is a spontaneous, nucleation-like event where a free energy barrier temporarily inhibits a thermodynamically more stable electron-lattice configuration, until a local fluctuation can overcome the barrier, triggering a collapse into a polaron state. Once the electron forms a polaron by creating a localized lattice deformation, an initial acoustic wavefront propagates outward from the deformation site, carrying away energy. This acoustic wavefront reflects the dynamic response of the lattice to the polaron formation. Following this initial wavefront, the system does not immediately settle into the exact many-body ground state but undergoes an oscillating relaxation process, emitting acoustic radiation. These outward-propagating oscillations form a radially oscillatory pattern, with wavelengths corresponding to the characteristic wavelength of the deformation potential, which closely matches the Debye wavelength—the maximum wavelength of acoustic modes in the lattice. Within the localized deformation, the electron remains bound but exhibits rattling motion. These oscillations drive the continued radiation of vibrations and eventually stabilize the lattice deformation into a Bessel-like equilibrium profile. The oscillation frequency inside the deformation is tied to the level spacing of the electron's quantized vibrational states within the polaron. The speed of the acoustic radiation is found to be slightly slower than the chosen nominal sound velocity. The time variations of the real-space averages of observables, such as position and mean square distance, resemble those of the static case; however, this behavior arises from self-trapping, where the electron's backaction on the lattice creates a classical-like potential well, lowering its energy and forming a bound state, unlike Anderson localization, which results purely from phase interference in a static disordered potential without any lattice backaction~\cite{PhysRevLett.47.882}. The apparent rotational asymmetry in the polaron state seen in Fig. 1g is a result of single-shot simulations with randomly initialized phonon phases run at a finite time. We verified the robustness of our results by averaging over 10 independent realizations and found no significant differences, confirming that the underlying physics remains unchanged. Note that here self-trapping refers to the formation of a strongly bound electron-lattice state due to non-perturbative interactions, rather than a sharp transition. All the while, the total energy remains constant up to numerical accuracy.

Several quantitative indicators confirm the formation of an acoustic polaron. Rapid oscillations in the electron autocorrelation function, beginning approximately 1 ps after the wavepacket launch, mark the onset of polaron formation. The time variations in electron kinetic energy, potential energy, and vibrational energy highlight the intense interaction between the electron and the lattice. In scenarios without back action, the electron’s kinetic and potential energies fluctuate around a few millielectronvolts, while the vibrational energy of the lattice remains constant. However, when the back action is included, the energies behave similarly until around 1 ps, after which the vibrational energy sharply increases to approximately 150 meV. This increase reflects the electron gaining kinetic energy from the dynamic lattice vibration field and the formation of a hole in the deformation potential, as indicated by the potential energy graph. The vibrational energy rise is a direct consequence of the strong back action of the electron on the lattice.

The total energy remains constant in both static and back action cases, as expected, but increases incorrectly if back action is neglected. With it included, energy conservation is restored, as it must be, because the total Hamiltonian is independent of time. Rapid oscillations in the electron time autocorrelation are further evidence of the formation of an acoustic polaron~\cite{doi:10.1143/JPSJ.65.1317}. The average momentum of the electron (not shown here) also exhibits similar fast oscillations, indicating the wiggling motion of the electron within the well it has created.  

It is a good time to herald the power of the coherent state representation: The entire scenes seen in Fig. \ref{fig1} and \ref{fig2}, including the polaron and waves emanating from it, and the thermal part of the deformation potential, plus information hidden in the snapshots about the rate of change of all these things, is {\it provided by a single multivariate Gaussian coherent state.} All that complexity is contained in one configuration.

The strong back action of the electron on the lattice leads to acoustic polaron formation at low temperatures, as demonstrated by the results for $T=10$ K shown in Fig. \ref{fig2}. The figures display a snapshot of the potential at time $t_f=3$ps, with the momentum space probability of the electron plotted in the bottom-left and the real-space probability in the bottom-right. For $m^*=10$ and $E_d=10$ eV, a single acoustic polaron forms. Doubling the effective mass results in the formation of a second, albeit weaker, polaron. When the deformation potential constant is doubled, multiple well polarons appear in close proximity to each other. Despite this, the multiple-well polaron configuration remains stable, with each polaron occupying its own distinct potential well. In fact, these correspond to quantum amplitudes for formation, which show up as separate events. There is no possibility of multiple polaron formation coming from one electron, except as amplitudes which must choose a singular realization upon measurement.

\begin{figure*}
\centering
  \includegraphics[width=0.8\textwidth]{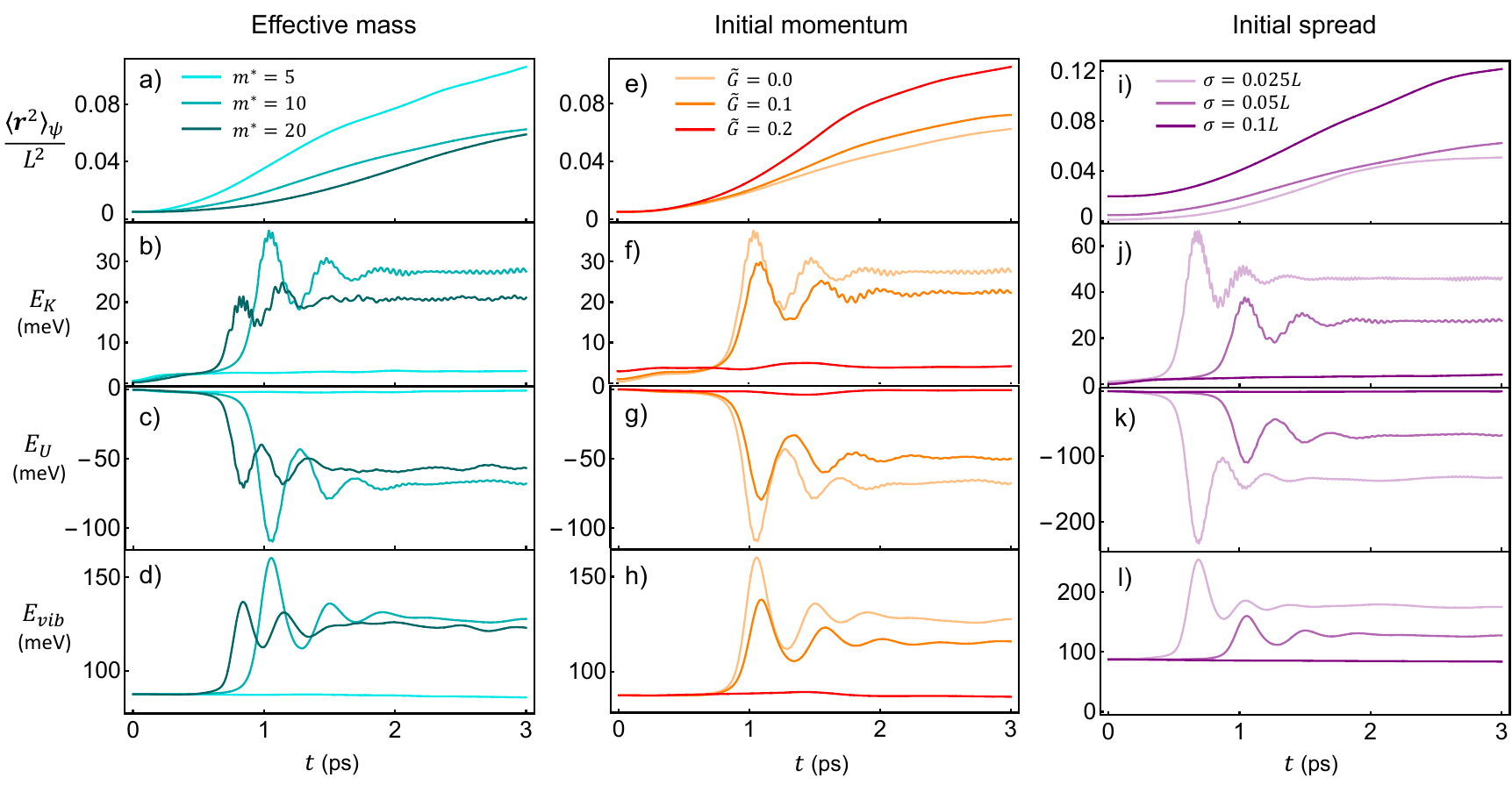}
  \caption{Time variations of the mean square distance, electron kinetic energy, potential energy, and vibrational energy are compared for different values of effective mass, initial velocity, and initial spread of the electron wavepacket. Acoustic polaron formation is more favorable with a high effective mass, low electron velocity, and a tightly confined electron wavepacket, with no external electric or magnetic fields applied.
  }\label{fig3}
\end{figure*}

\begin{figure*}
\centering
  \includegraphics[width=0.8\textwidth]{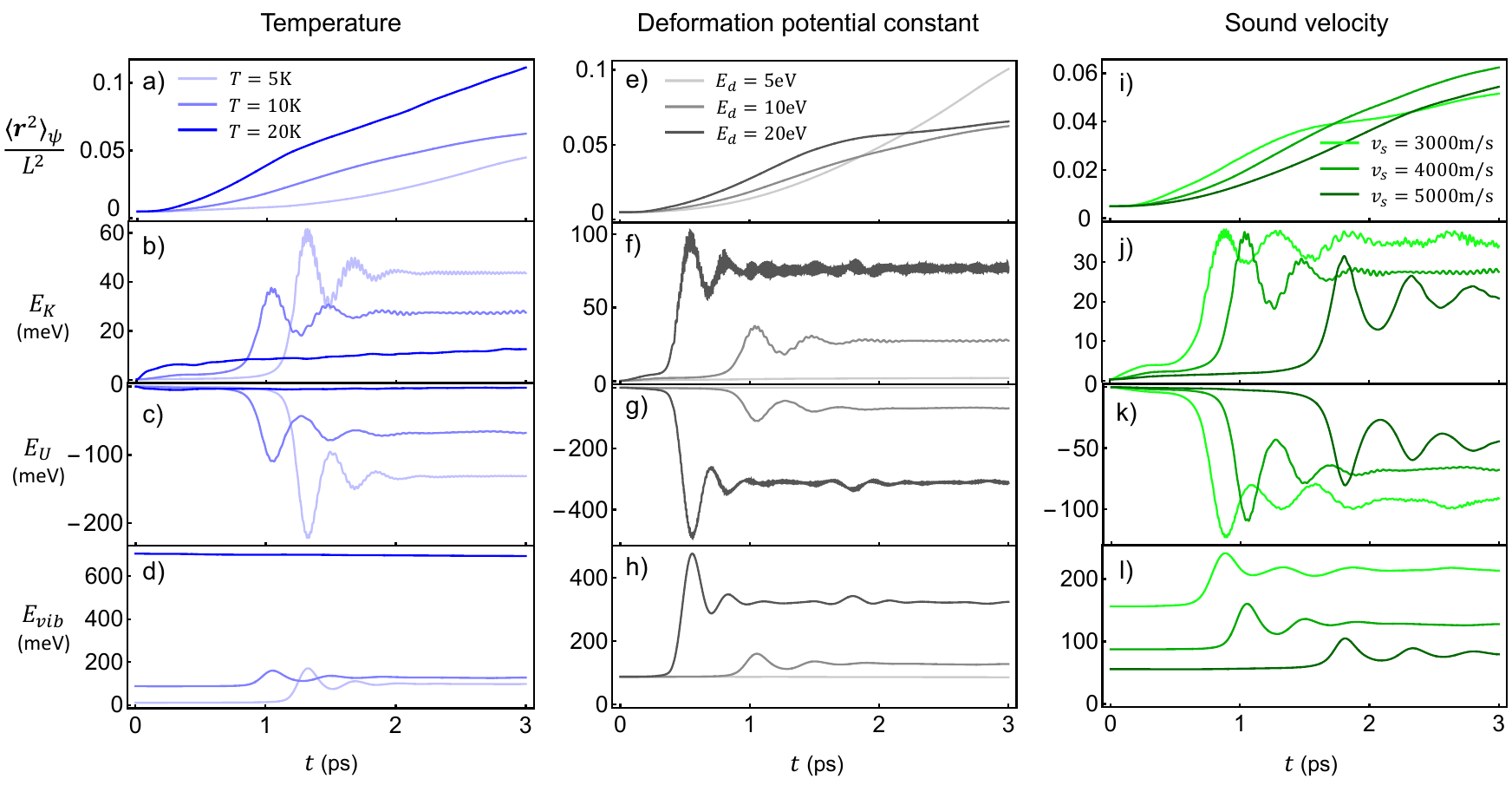}
  \caption{Time variations of the mean square distance, electron kinetic energy, potential energy, and vibrational energy are compared for different values of temperature, deformation potential constant, and sound velocity. Acoustic polaron formation is easier at low temperatures, high deformation potential constants, and slow sound velocities, with no external electric or magnetic fields applied.
  }\label{fig4}
\end{figure*}

\section{Analysis of polaron formation}\label{s:sec4}

In this section, we examine the ease or difficulty of polaron formation in relation to various material and system parameters, as depicted in Figs. 3 and 4. The time evolution of the mean square distance provides a measure of the system’s real-space dynamics. We focus on the time dependence of the total and sub-energies. We compare polaron formation under three distinct values for each parameter, relative to the reference case which is shown in Fig. \ref{fig1}.

The binding energy of the polaron is the difference between the free electron’s energy before forming the polaron (i.e. unbound energy) and the electron kinetic energy plus the potential energy in the well (i.e. the total energy of the bound electron state), $E_{\text{b}}=E_{\text{unbound}}-E_{\text{bound}}$, where $E_{\text{unbound}}=E_{\text{K}}(0)+E_{\text{U}}(0)$, and $E_{\text{bound}}=E_{\text{K}}(t_f)+E_{\text{U}}(t_f)$.

\begin{table}[]
\caption{Acoustic polaron binding energies (meV) across various cases. Entries are omitted where polaron formation was not observed.}
\label{tab:my-table}
\begin{tabular}{llc}
\multicolumn{2}{l}{Parameters and their values}                    & Binding energy (meV) \\ \hline
\multirow{3}{*}{Effective mass}   & $m^{*}=5~m_e$              & -    \\
                                  & $m^{*}=10~m_e$             & 4.07 \\
                                  & $m^{*}=20~m_e$             & 3.57 \\ \midrule
\multirow{3}{*}{Initial momentum} & $G=0$                     & 4.07 \\
                                  & $G=0.1$                   & 2.89 \\
                                  & $G=0.2$                   & -    \\ \midrule
\multirow{3}{*}{Initial spread}   & $\sigma=0.025$~L           & 8.79 \\
                                  & $\sigma=0.05$~L            & 4.07 \\
                                  & $\sigma=0.1$~L             & -    \\ \midrule
\multirow{3}{*}{Temperature}      & $T=5$~K                    & 8.77 \\
                                  & $T=10$~K                   & 4.07 \\
                                  & $T=20$~K                   & -    \\ \midrule
\multirow{3}{*}{Def. pot. const.} & $E_d=5$~eV                 & -    \\
                                  & $E_d=10$~eV                & 4.07 \\
                                  & $E_d=20$~eV                & 23.6 \\ \midrule
\multirow{3}{*}{Sound velocity}   & $v_s=3000$~m/s             & 5.76 \\
                                  & $v_s=4000$~m/s             & 4.07 \\
                                  & $v_s=5000$~m/s             & 2.41 \\ \midrule
\multirow{3}{*}{Electric field}   & $\vec{E}=0$               & 4.07 \\
                                  & $\vec{E}=10^5$~V/m         & 3.93 \\
                                  & $\vec{E}=5\times 10^5$~V/m & 2.40 \\ \midrule
\multirow{3}{*}{Magnetic field}   & $\vec{B}=0$               & 4.07 \\
                                  & $\vec{B}=10$~T             & 3.91 \\
                                  & $\vec{B}=50$~T             & 3.06 \\ \bottomrule
\end{tabular}
\end{table}

Fig. \ref{fig3} presents the results for varying effective mass, initial wavepacket velocity, and initial wavepacket spread. An increased effective mass lowers electron mobility and, thus, allows the electron more time to distort the lattice. Essentially, a heavier electron can more effectively “drag” the lattice along with it, facilitating the self-trapping process that characterizes polaron formation.

The initial velocity of the electron plays a critical role in the formation of polarons. A slower initial velocity allows the electron to spend more time interacting with a given region of the lattice,  increasing the likelihood of self-trapping. With less kinetic energy, the electron is more easily captured by the deformation potential created by the lattice vibrations, leading to the formation of a stable polaron.   At higher initial velocities, the electron moves too quickly for a stable deformation potential to form around it. Polaron formation is not observed at higher velocities. 

A more localized electron wavepacket (smaller spread) means that the electron’s probability density is concentrated in a smaller region of space. This increased localization enhances the electron’s interaction with the lattice, making it easier and faster for the electron to distort the lattice and form a polaron. The concentrated interaction facilitates the self-trapping process, leading to earlier polaron formation. A large wavepacket spread dilutes the interaction with the lattice. However, these are to some extent consequences of the mean-field approximation. Under full quantum dynamics, there are amplitudes for the electron to have been fully present at different locations, and under wavefunction collapse (or decoherence) the realization of one of these amplitudes associates one electron to the polaron. In the mean field, polaron size and energy depend upon the local value of $\vert \psi\vert^2$.

During the self-trapping process of the polaron, under the chosen initial parameters, the integrated probability density within a 1.5 nm radius ring around the polaron center is approximately 40\% at the moment of formation. As the process continues, the surrounding probability density is drawn into the polaron, eventually settling at approximately 60\% within the ring.

To estimate the mass of the formed polaron, we treat its confinement as a harmonic oscillator, thus determining $m_p = k / \omega^2$, where $k$ is the effective trapping stiffness and $\omega$ is the characteristic frequency of oscillations observed in our simulations. From the second derivative of the polaron potential, we determined $k \approx 0.44$ J/m², while the extracted oscillation frequency was $\omega \approx 1.23 \times 10^{14}$ Hz. This estimation yields a polaron mass of approximately 30 times the bare electron mass, which is consistent with reported values for acoustic polarons in the literature~\cite{PhysRevB.104.235123,PhysRevB.93.144302,polaronreview}.

After the formation of a well-defined polaron, its stability is evaluated by introducing a secondary wavepacket into the preformed deformation potential. At 3 ps, when the polaron has reached a fixed geometry, the first electronic wavefunction is substituted with a newly launched wavepacket. This second wavepacket is centered inside the polaron with one-fourth of the original spread. The analysis reveals that even after 6 ps, the polaron retains its spatial structure, with $> 99\%$ of the probability amplitude remaining localized within the preformed region, thereby confirming its high stability.

In Fig. \ref{fig4}, we present the results for various temperatures, deformation potential constants, and sound velocities. At lower temperatures, lattice vibrations are less energetic, which reduces the disruptive scattering processes that could otherwise prevent the electron from forming a stable polaron. Consequently, the electron can more effectively interact with the lattice, leading to stronger binding energies and more pronounced polaron formation. At 5K, for example, larger polaron binding energy is observed due to less disruption from lattice vibrations. On the other hand, at 10K, the electron forms a polaron more quickly because it finds an optimal trapping site faster, aided by the potential landscape at that temperature. Notably, polaron formation is not observed above 20K, due to strong and dynamic deformation potential scattering. Additionally, we examined the stability of the polaron under gradual temperature increases after its formation. Our findings indicate that the polaron remains robust, retaining its structure even at elevated temperatures, provided the temperature rise is not extreme.

The deformation potential constant, $E_d$, directly influences the strength of the electron-phonon interaction. A higher $E_d$ enhances the coupling between the electron and the lattice vibrations, leading to earlier polaron formation and greater binding energy. Specifically, for $E_d=20$~eV, the strong interaction causes the polaron to form rapidly with very high binding energy. At $E_d=10$~eV, the formation is slower, with moderately lower binding energy, while at $E_d=5$~eV, the interaction is too weak to sustain polaron formation, resulting in no observable polaron.

\begin{figure}
\centering
  \includegraphics[width=0.49\textwidth]{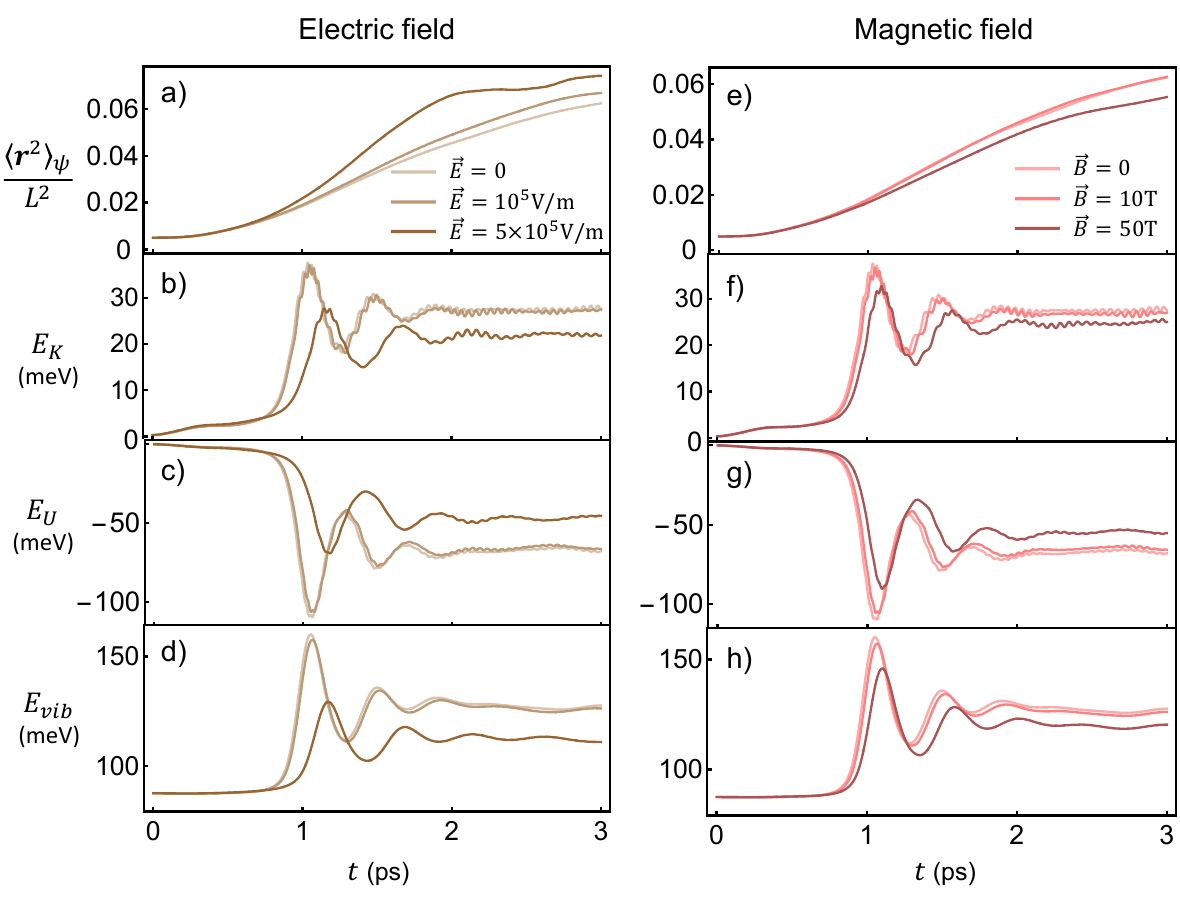}
  \caption{Time variations of the mean square distance, electron kinetic energy, potential energy, and vibrational energy in the presence of electric and magnetic fields are compared for their different values. Rising electric and magnetic fields hinder the formation of acoustic polarons.
  }\label{fig5}
\end{figure}

Sound velocity, $v_s$ determines the propagation speed of lattice vibrations. Here it corresponds to how fast the deformation potential moves in time. When the sound velocity is slower, the deformation potential changes more gradually, allowing the electron more time to interact and self-trap within this potential landscape. With a slower-moving deformation potential, the electron experiences less scattering and disruption, making it easier for the electron to localize and form a polaron. The electron can more effectively establish a stable interaction with the lattice, facilitating earlier polaron formation. Even though slower sound velocities favor earlier polaron formation, within the typical range of sound velocities found in materials, the deformation potential still moves slowly enough to allow for polaron formation. While faster sound velocities might introduce more scattering and disruption, they are generally not fast enough to entirely prevent polaron formation, ensuring that the polaron can still consistently form.

The results with electric and magnetic fields applied are shown in Fig. \ref{fig5}. Both electric and magnetic fields introduce additional forces that can disrupt the delicate balance required for polaron formation. These fields can alter the electron’s trajectory and reduce the effectiveness of its interaction with the lattice, leading to a decrease in the binding energy of the polaron. Consequently, the presence of these fields delays the formation of the polaron as the electron takes longer to settle into a stable, self-trapped state. Despite these effects, the overall impact of electric and magnetic fields on polaron dynamics remains relatively minor within the range of reasonable field strengths. This suggests that while these fields can influence polaron formation, they do not significantly alter the fundamental process unless the field strengths are exceptionally high.

\section{Discussion}\label{s:sec5}
In this study, we have used the quantum acoustic framework to demonstrate acoustic polaron formation arising from non-perturbative and coherent scattering processes between electrons and lattice vibrations. We have derived and solved the coupled equations of motion for the electron wavepacket and lattice vibration coherent state, and studied the dynamics of the coupled system in real time and real space. We have determined the acoustic polaron binding energies.  We have established the conditions necessary for the formation of acoustic polarons, for a wide range of material parameters.  We provided a comprehensive study on the effect of various material parameters on the formation and stability of the acoustic polarons. 

The quasi-classical field approach we use here, while computationally efficient and capable of capturing key dynamics, does have certain limitations~\cite{li2013polaron}. One such limitation is the lack of explicit accounting for quantum correlations and entanglement between the electron and the lattice. This omission means that interference between amplitudes for events occurring at different locations and times is not fully represented. However, such interference is unlikely to significantly affect bulk properties, such as the spatial extent of electron motion, particularly considering that the polaron amplitudes in different locations are expected to decohere. This decoherence is naturally implied by the lattice wave emissions observed in Fig.~\ref{fig2}.

Another limitation lies in the dependence of the lattice distortion on the local electron density. This dependence can introduce some sensitivity to the initial wavepacket spread. Nevertheless, as highlighted in Fig.~\ref{fig3}, we demonstrate that the observed polaron energies and dynamics are fairly robust, with modest effects arising from variations in the initial wavepacket. While deeper polaron binding might be expected for higher local electron densities, the overall trends captured here provide reliable insights into the mechanisms of polaron formation and the influence of key material parameters.

The backaction reveals a pathway for studying intricate physical phenomena, such as charge density waves, which emerge when the electron density couples strongly to specific phonon modes, and the Kohn anomaly, characterized by a softening of phonon frequencies near the Fermi surface due to enhanced electron-phonon coupling. Hence, the present paper is the first of a series that we expect will follow, with back-action included. We expect to be able to examine bound bipolarons by similar techniques.

\section*{Acknowledgments}
We thank Supriyo Datta, Joonho Lee, Joost de Nijs, Nikolai Leopold, and Donghwan Kim for the useful discussions. We thank Zunqi Li, Hongkun Chen, for their contributions to discussions during their internship. This work was supported by the U.S. Department of Energy under Grant No. DE-SC0025489. A.A. acknowledges financial support from the Sabanci University President’s Research Grant with project code F.A.CF.24-02932. A.M.G. thanks the Studienstiftung des Deutschen Volkes for financial support. J.K.-R. thanks the Oskar Huttunen Foundation for the financial support.

\bibliography{refs}
\bibliographystyle{unsrt}
\end{document}